\documentclass[final]{revtex4}
\usepackage{graphicx}
\usepackage{amssymb}
\usepackage{hyperref}

\newcommand{\be}{\begin{equation}}
\newcommand{\ee}{\end{equation}}
\newcommand{\bea}{\begin{eqnarray}}
\newcommand{\eea}{\end{eqnarray}}

\newcommand{\half}{\frac{1}{2}}

\newcommand{\etal}{{\it et al.},}

\newcommand{\ba}{\begin{array}}
\newcommand{\ea}{\end{array}}
\newcommand{\bi}{\begin{itemize}}
\newcommand{\ei}{\end{itemize}}
\newcommand{\ben}{\begin{enumerate}}
\newcommand{\een}{\end{enumerate}}

\begin{document}
\title{\vspace*{.5in}Dark matter and pulsar signals for \\ Fermi LAT, PAMELA, ATIC, HESS and WMAP data}
\author{V.~Barger$^{1}$, Y.~Gao$^1$, W.-Y. Keung$^2$, D.~Marfatia$^{3}$, G. Shaughnessy$^{4,5}$}
\affiliation{$^1$Department of Physics, University of Wisconsin, Madison, WI 53706}
\affiliation{$^2$Department of Physics, University of Illinois, Chicago, IL 60607}
\affiliation{$^3$Department of Physics and Astronomy, University of Kansas, Lawrence, KS 66045}
\affiliation{$^4$Department of Physics and Astronomy, Northwestern University, Evanston, IL 60208}
\affiliation{$^5$High Energy Physics Division, Argonne National Laboratory, Argonne, IL 60439}
\begin{abstract}
We analyze new diffuse gamma-ray data from the Fermi Gamma-ray Space Telescope, which do not confirm an excess in the EGRET data at 
galactic mid-latitudes, in combination with measurements of electron and positron fluxes 
from PAMELA, Fermi and HESS within the context of three possible sources: dark matter (DM) annihilation or decay into charged leptons, 
and a continuum distribution of pulsars. 
We allow for variations in the backgrounds, consider several DM halo profiles, 
and account for systematic uncertainties in data where possible. We find that all three scenarios represent the data well. The pulsar
description holds for a wide range of injection energy spectra. We compare with ATIC data and the WMAP haze where appropriate, but do not 
fit these data since the former are discrepant with Fermi data and the latter are subject to large systematic uncertainties. 
We show that for cusped halo profiles, Fermi could observe
a spectacular gamma-ray signal of DM annihilation from the galactic center
while seeing no excess at mid-latitudes.
\end{abstract}
\thispagestyle{empty}
\maketitle
{\bf Introduction.}
Recent observations of an excess in cosmic ray positrons have led to much excitement in the particle physics community.  
At the heart of this excitement is the possibility that these may be a signal of the annihilation or decay of cold dark matter (DM) 
- the nonbaryonic, nonluminous component of matter that constitutes 20\% of the energy budget of the universe, and is either exactly stable
or quasi-stable over a time-scale that is much longer than the age of the Universe. 

The Payload for Antimatter Matter Exploration and Light-nuclei Astrophysics (PAMELA) experiment~\cite{Adriani:2008zr} reports a prominent 
upturn in the positron fraction 
from $10-100$ GeV, in contrast to
what is expected from high-energy cosmic rays interacting with the
interstellar medium. (We employ the updated PAMELA spectrum~\cite{pamnew}, derived using 2.5 times more data and refined background 
rejection methods, and which is slightly softer than the published spectrum in 
Ref.~\cite{Adriani:2008zr}.)
This result confirms excesses seen in
previous experiments, such as High Energy Antimatter Telescope (HEAT)~\cite{HEAT}
and Alpha Magnetic Spectrometer (AMS-01)~\cite{Aguilar:2007yf}. 
The PAMELA satellite has a calorimeter and magnetic spectrometer calibrated up to 200~GeV and requires
discrimination between positrons from protons at the level of $2\times 10^{-4}$. Due to its small size, a large fraction of the 
electromagnetic shower from $e^{\pm}$ is not contained, making the separation from protons and antiprotons (which do not shower) difficult.
If the data are taken at face value, the rise in $e^+$ fraction is indicative of some source that injects high energy positrons that are not included 
in the standard astrophysical background estimates.  PAMELA did not find an excess of antiprotons~\cite{Adriani:2008zq}, which may be indicative of a 
leptophilic DM particle~\cite{dm-leptophilic} or pulsars~\cite{pulsar,profumo,Malyshev:2009tw}.  Data obtained by the Advanced Thin Ionization Calorimeter (ATIC)~\cite{Chang:2008aa} and the Polar Patrol Balloon and Balloon borne Electron Telescope with Scintillating fibers (PPB-BETS)~\cite{Torii:2008xu} experiments show an excess in the $e^++e^-$ flux from 
200 to 800 GeV. The challenge for these experiments, which can not discriminate charge, is to distinguish $e^{\pm}$ from neutral pions at the level 
of $10^{-3}$. The excess seen by these experiments is in an uncalibrated energy regime above 200~GeV.
It is worth mentioning that ATIC has also found bumps in the spectra of light cosmic nuclei that remain puzzling~\cite{lightnucleibump}. 
Preliminary results from ATIC-4, which has a larger
calorimeter to contain the electromagnetic showers, confirm their previous data.
In addition, the High Energy Stereoscopic System (HESS) $\gamma-$ray telescope has inferred a 
flat but statistically limited $e^++e^-$ spectrum between 340~GeV and 1~TeV~\cite{hesslow} which falls steeply above 
1~TeV~\cite{Collaboration:2008aaa} indicating a high energy limit of the new $e^\pm$ source. 
The extraction of the $e^\pm$ spectrum, which is the first of its kind, 
is made difficult by contamination from known $\gamma-$ray 
sources and the hadronic cosmic ray background. Although designed as a $\gamma-$ray observatory, 
the Large Area Telescope (LAT) of the Fermi Gamma-ray Space Telescope~\cite{Baltz:2008wd} can also detect
$e^\pm$ without charge discrimination. Recent high-statistics data from Fermi show an almost flat $e^++e^-$ spectrum between 20~GeV and 1~TeV,
with a slight excess between 200~GeV and 1~TeV, but no bump-like structure~\cite{fermilat}, in contradiction with ATIC.
It has been noted that a slightly harder supernova injection spectrum for the electron background can reproduce the Fermi data (at the 
expense of agreement with lower energy data), but can not account for the PAMELA excess~\cite{grasso}.

%

Besides direct observations of excessive cosmic $e^\pm$, the Energetic Gamma Ray Experiment Telescope (EGRET) observations 
provide hints of an excess of diffuse $\gamma-$rays from $10-50$~GeV.
Reanalysis of the data using improved sensitivity estimates~\cite{Thompson:2004ez} 
found a harder spectrum at these energies~\cite{Strong:2005zx} than previously derived.
 The Fermi LAT observes $\gamma-$rays from 
20 MeV to 300 GeV. Preliminary results from Fermi~\cite{glastpri} show no readily discernible excess above backgrounds in the GeV  range. 

The  Wilkinson Microwave Anisotropy Probe (WMAP) observes microwave radiation in the 23 to 94 GHz range. 
A residual microwave background has been identified in WMAP data
after known astrophysical backgrounds are subtracted from the skymap~\cite{finkbeiner,Cumberbatch:2009ji}. 
This haze seems to originate from the synchrotron radiation of high energy electrons that may be attributable to 
annihilations of a thermal relic particle~\cite{finkbeiner2}. However, the systematic uncertainty of the WMAP haze has not been 
robustly established, especially in the inner galactic region. As pointed out in Ref.~\cite{Cumberbatch:2009ji}, the significance
of the haze can be substantially reduced by allowing a spatial variation in the frequency dependence of the synchrotron radiation
in the inner and outer galactic regions.

In this letter, we perform a comparative study of three scenarios (described in the next section) that have been proposed to explain the 
aforementioned data. We do not include ATIC data in our statistical analyses since they are not corroborated by Fermi and we do not include
the low energy HESS data sample~\cite{hesslow} since the Fermi dataset is significantly larger in the overlapping energy region below 1~TeV.
Nevertheless, we comment on ATIC data throughout the text.
A distinguishing feature of our analysis is that we vary the power-law index for the electron background injection spectrum, account
for cosmic ray propagation uncertainties and energy calibration uncertainties in data (which can rigidly shift spectra significantly), and consider several DM halo profiles. All these have the potential to alter conclusions drawn from less thorough analyses.

{\bf Sources.}
{{\it (i) Annihilating Dark Matter.}}
\label{anni}
Genuinely stable DM, with a discrete symmetry that enforces stability, is central to many scenarios 
such as Supersymmetry, minimal Universal Extra Dimensions and Little Higgs with T-parity.  
Typically, the DM candidate interacts weakly and has a mass $M_{DM}$ in the $100-1000$~GeV range.  
Throughout, we set the nonrelativistic thermally averaged annihilation cross section to the canonical value 
$\langle \sigma  v\rangle = 3\times 10^{-26}$ cm$^3$s$^{-1}$,
required for  saturating the DM relic density in the standard cosmological scenario.
Direct annihilation or fragmentation to antiparticle states or high energy gamma-rays may yield
a contribution above the astrophysical background. For recent studies related to the PAMELA and ATIC data see Refs.~\cite{bib:bkmg,cirelli,Huh:2008vj}.

{{\it (ii) Decaying Dark Matter.}}
\label{decay}
If the exact discrete symmetry that protects the stability of the DM particle is slightly broken, DM may decay on time-scales longer than 
the age of the universe. Phenomenologically, this case is similar to DM annihilation with the exception that the DM mass 
is a factor of two larger, and the injection rate depends linearly (instead of quadratically) on the density of DM at the source. 
See Ref.~\cite{dm-dec} for recent work.

{{\it (iii) Pulsars.}}
\label{pulsar}
An alternative explanation of the PAMELA and ATIC excess is pulsars that emit high energy positrons~\cite{pulsar,profumo,Malyshev:2009tw}. 
Electrons in the intense rotating 
magnetic field that surrounds the neutron star can emit synchrotron radiation that is energetic enough to produce electron and positron pairs.  
The resulting positrons can propagate away from the pulsar via open magnetic field lines or via the pulsar wind.  
The resulting positron spectra can be modeled as a product of a power law and a decaying exponential with a cutoff at energy $E_p$.

To conveniently present our results in terms of their dependence on a generic energy scale, $E_s$, of injected 
positrons for the three cases above, we define
\be
E_{s} \equiv \left\{
\begin{array}{cc}
M_{DM} & \text{Annihilating DM}\\
\half M_{DM}& \text{Decaying DM}\\
E_p& \text{Pulsars}\\
\end{array}
\right. .
\ee

\begin{figure}
\begin{center}
\includegraphics[scale=0.62]{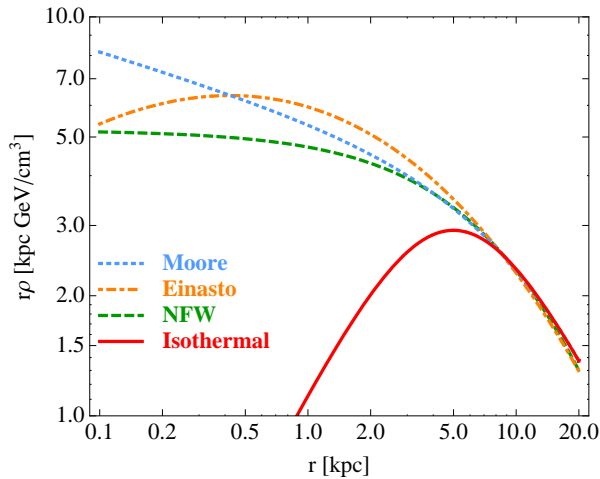}
\caption{$r\rho(r)$ for the Moore, Einasto, NFW, and isothermal profiles
with density 0.3~GeV/cm$^3$ at 8.5 kpc. 
The analytical expressions are provided in the appendix.}
\label{fig:profileplot}
\end{center}
\end{figure}

{\bf Cosmic ray species.}
\label{CR}
Numerous astrophysical sources produce high energy cosmic rays that propagate through turbulent galactic magnetic fields.  
We utilize the GALPROP code~\cite{Strong:1999sv} to simulate this propagation. For the distribution of the DM halo 
we consider the isothermal~\cite{IsoT}, Navarro, Frenk \& White (NFW)~\cite{Navarro:1995iw}, Einasto~\cite{Navarro:2008kc} and 
Moore~\cite{Diemand:2005wv} profiles; see Fig.~\ref{fig:profileplot}. 


{\underline {\it Antimatter.}}
\label{posi}
Of the various cosmic ray species, antimatter provides one of the best avenues for probing DM because of relatively low backgrounds.  
The majority of the galactic $e^+$ background is created by the decay of cosmic rays or their interaction with the interstellar medium.  

{\it (i) Annihilating Dark Matter.}
The positron spectrum from DM annihilation strongly depends on the underlying model and spin of the DM particle.  
To explain the hard positron spectra observed by PAMELA, a dominant annihilation fraction to charged leptons is preferred~\cite{bib:bkmg,cirelli}. 
For the excess in the ATIC data, $M_{DM}$ between $600-1000$~GeV is required.  
However, the cross section needed to 
reproduce the relic density is about two orders of magnitude lower than that required to fit the $e^\pm$ data.  
This has led many to invoke a 
boost factor from $s$-channel resonances~\cite{Ibe:2008ye}, the nonrelativistic enhancement from the Sommerfeld effect~\cite{bib:sommerfeld}, or
DM clumping in the halo substructure~\cite{dm-clumping}.  A boost factor of astrophysical origin is easily motivated since local clumps in the 
halo distribution can enhance the flux from DM annihilation. N-body simulations suggest boost factors of order 10~\cite{Lavalle:1900wn}, 
but it must be borne in mind that 
the simulations neglect ordinary matter. In principle, the flux from $e^{\pm}$, antiprotons and photons may have different boost factors that are 
energy dependent, and depend on astrophysical assumptions such as the inner density profiles of clumps.  
We assume equal energy-independent boost factors for all species. 

In contrast to their positron data, PAMELA observes no excess in the antiproton spectrum and therefore rules out most processes 
that produce $e^+$ and hardonic states at roughly the same rate.  For instance $W, Z$ and quark pair productions generally produce 
too many antiprotons to agree with both the $e^+$ and $\bar p$ PAMELA data~\cite{bib:bkmg,cirelli}. 
However, if the DM particles annihilate dominantly into leptons, consistency with the antiproton data
can be achieved even if the DM mass is a few hundred GeV due to the suppressed fragmentation to $\bar p$.   
We consider cases in which DM annihilates to $e^+e^-$, $\mu^+\mu^-$ and $\tau^+\tau^-$ with 100\% branching fractions, and to these channels with equal branching fractions.  

{\it (ii) Decaying Dark Matter.}
 For fermion DM, decay channels such as $W^{\pm}\ell^{\mp}$ with $\ell$ 
a lepton, are possible, and the $e^+$ spectrum originates from equally weighted $\ell^+$ and $W$ fragmentation. However, the associated $W^\pm$ may 
contribute too many antiprotons to be consistent with PAMELA.  Also, the 3-body decay mode $\ell^+\ell^-\nu$ is permissible, 
but model-dependent.  It is noteworthy that 
decay modes that are helicity suppressed are not intrinsically disfavored since the lifetime suggested by the data is significantly long. 
 For the sake of direct comparison with DM annihilation, we restrict our study to 2-body decay modes into lepton pairs with equal branching fractions. 
Thus, for DM decay we effectively assume that the DM particle is a boson. 

A nice feature of DM decay is that no boost factors are required since the DM relic abundance is produced via annihilation while
 the astrophysical signals arise from DM decay.

{\it (iii) Pulsars.}
It is unlikely that a single pulsar is responsible for both the PAMELA and ATIC results. The PAMELA data favor a mature pulsar (2~Myr) 
at a distance of a kpc, or a younger and closer pulsar like Geminga (0.37 Myr old at 160~pc), while the ATIC data require a very powerful and younger (0.5~Myr) pulsar
at a distance of $1-2$ kpc~\cite{profumo}. Rather than considering known nearby pulsars, we 
follow Ref.~\cite{zhangli} and model the pulsar sources as a continuum distribution of pulsars throughout the galaxy in cylindrical coordinates as 
\be
\rho({\tt r}) = N\, \left(\frac{\tt r}{{\tt r}_{\odot}}\right)^a e^{-\frac{b(\tt{r}-\tt{r}_{\odot})}{\tt{r}_{\odot}}}
e^{-\frac{z}{z_0}}\,,
\label{pulsardist}
\ee
where $z_0=0.2$~kpc, $\tt{r}$ is the cylindrical radius from the galactic center parallel to the galactic disk, and ${\tt r}_{\odot}=8.5$~kpc.   
For mature pulsars, we take $a=1.0$ and $b=1.8$~\cite{zhangli}. 
$N$ is a normalization that is determined by fitting PAMELA, Fermi and HESS data.

We take the positron injection spectrum to be
\be
\frac{dN_{e^{\pm}}}{dE}\varpropto E^{-\alpha}e^{-E/E_{p}}\,, 
\label{eq:pulsar}
\ee
where we fix $\alpha=1.5$ and allow $E_p$ to span the range from 150~GeV to 2~TeV. By fixing $\alpha$, the number of independent parameters
that describe the combined signal and background spectra for pulsars is the same as that for DM annihilations and decays. Note that
variations in $\alpha$ can be somewhat compensated by $N$, thus making our results only mildly sensitive to $\alpha$ over acceptable ranges.
However, it should be noted that pulsars of various ages may well possess a continuum of cut-off energies~\cite{Malyshev:2009tw} and an 
age distribution may lead to nontrivial structures in the high energy spectrum, with young pulsars contributing more dominantly to the 
high energy range than mature pulsars. 

Pulsars do not produce antiprotons because of the absence of hardonic showers, and thus naturally explain why the PAMELA antiproton spectrum matches the expected spectrum.

{\underline {\it Light nuclei.}}
\label{lightnuclei}
Although not as compelling as the positron excess, the ATIC experiment has found excesses in the spectra of several light nuclei 
including C, N, O and Si, in the same energy range as the $e^++e^-$ excess~\cite{lightnucleibump}.
None of the scenarios under consideration offer an explanation of these anomalies, and we do not consider them any further.


{\underline {\it Photons.}}
\label{photons}
Under our assumption that the DM and pulsar sources produce only charged leptons directly, the main contribution to the flux of high energy photons is
 final state radiation (FSR) from the $e^+e^-$ mode and secondary decays.  
In addition, there is a continuum at lower energies from bremsstrahlung as the leptons leave the source. 
Once the leptons are ejected into the galactic environment, they interact with the galactic magnetic field and the interstellar
radiation field (ISRF). They lose energy via inverse compton (IC) scattering which produces gamma radiation and via synchroton radiation as radiowaves.  
The diffuse gamma-ray background can be modeled as a sum of galactic pion decay (which dominates in the $1-10$~GeV range), 
IC scattering, bremsstrahlung and the isotropic extragalactic (EG)~\cite{Strong:2004ry} diffuse gamma-rays (estimated up to about 50~GeV and which become
increasingly relevant for higher latitudes). 


\begin{figure*}[b]
\begin{center}
\includegraphics[scale=0.57]{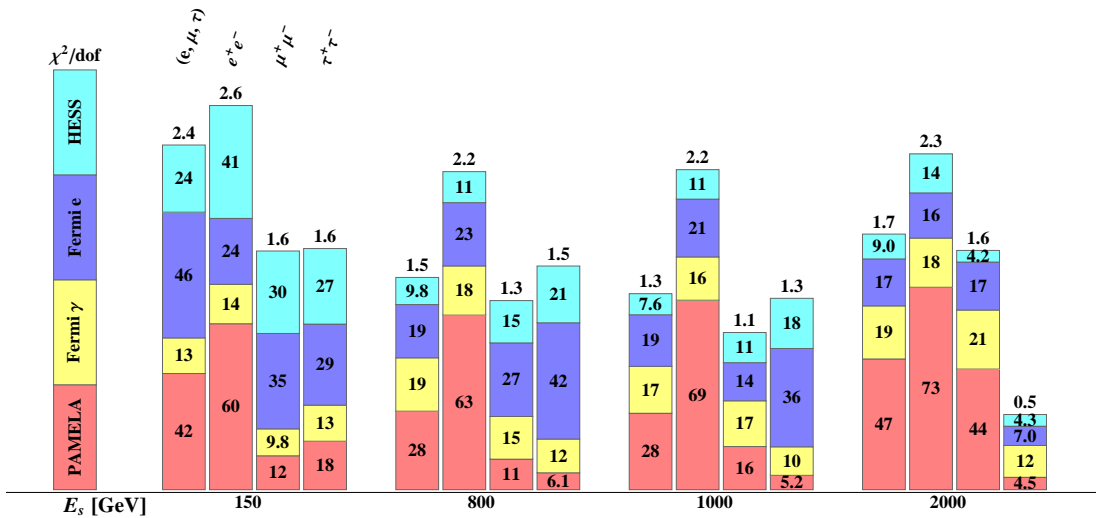}
\caption{Best-fit $\chi^2$ values for DM annihilation into various charged lepton final states 
for several values of $E_s$ from a joint analysis of the PAMELA, Fermi $\gamma-$ray, Fermi $e^\pm$ and HESS datasets 
which have 7, 18, 26 and 8 points, respectively. $(e,\mu,\tau)$ denotes a democratic final state with equal branching fractions into
$e^+e^-$, $\mu^+\mu^-$ and $\tau^+\tau^-$.
The number of free parameters is 8, and the number of degrees of freedom (dof) is 53, including two energy scale normalizations; 
see the appendix for details. The $\chi^2$/dof is provided above each column.
For $E_s=150$~GeV, all data other than the falling HESS spectrum are well reproduced for the $\mu^+\mu^-$ and $\tau^+\tau^-$ modes.
An isothermal halo profile is assumed. 
 }
\label{fig:baranni}
\end{center}
\end{figure*}

\begin{figure*}[t]
\begin{center}
\includegraphics[scale=0.7]{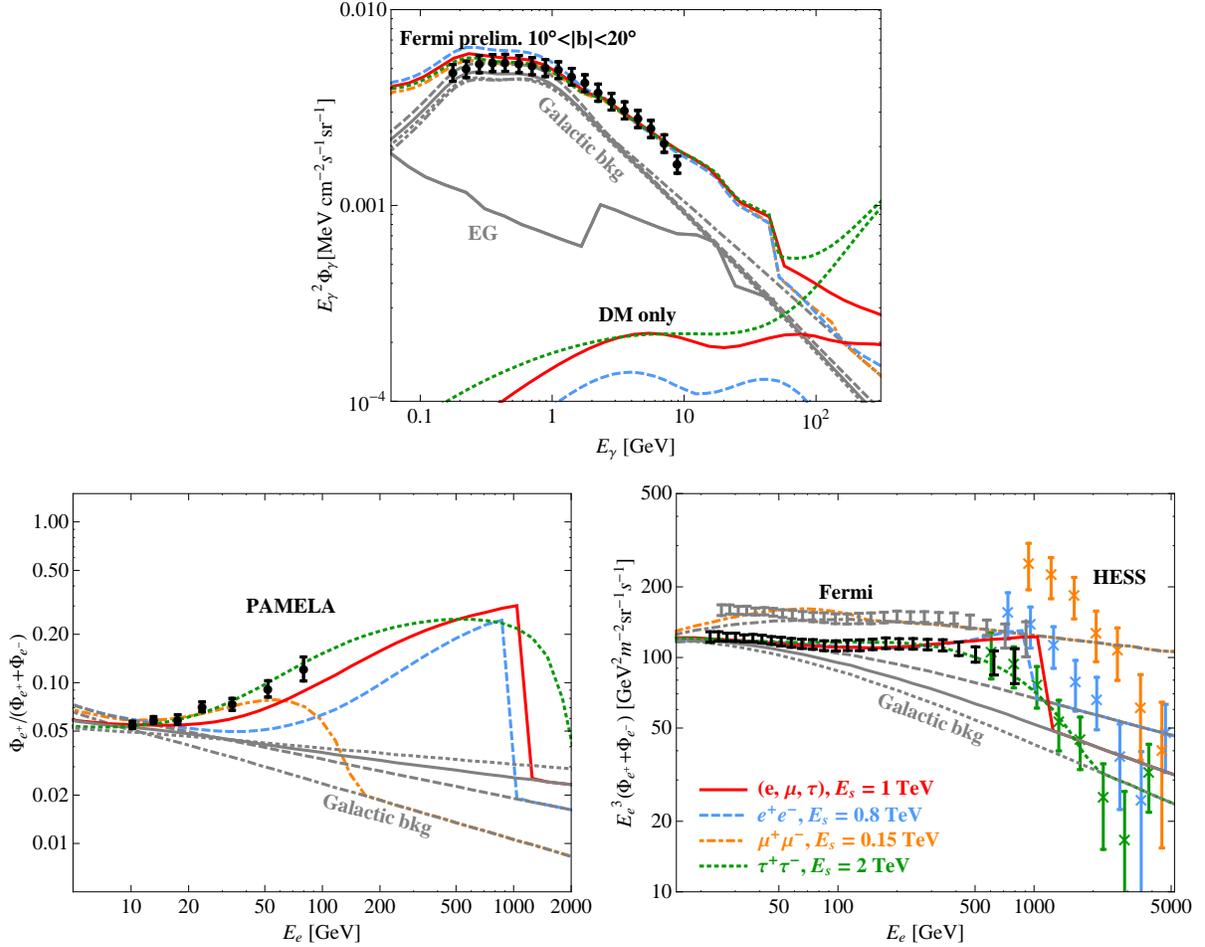}
\caption{Illustrative spectra for DM annihilation with $E_s=150$~GeV, 800~GeV, 1~TeV and 2~TeV into $\mu^+\mu^-$, $e^+e^-$, $(e,\mu,\tau)$, and 
$\tau^+\tau^-$, respectively. The DM-only contribution to $\gamma-$rays for the $\mu^+\mu^-$ mode is tiny (because the boost factor is about 25; see 
Eq.~\ref{bf}) and lies outside the range of the figure. The corresponding
$\chi^2$ values can be found in Fig.~\ref{fig:baranni}. The background contribution for each case is displayed in the same line-type as for the signal. 
From top to bottom in the bottom-right panel, the electron background injection spectral indices are $\gamma_0=2.38$, 2.49, 2.57 and 2.63; 
see the appendix. The orange, light blue and dark green HESS data points (color coded according to the annihilation channel) 
are shifted within the energy calibration uncertainty with $\epsilon=1.3$, 1.02 and 0.84, 
respectively. The HESS data corresponding to annihilation into $(e,\mu,\tau)$ have $\epsilon=0.98$, and are not shown.
For $E_s=800$~GeV, 1~TeV and 2~TeV, the Fermi $e^\pm$ data shift downward (shown in black) with an energy calibration scale of 0.9, and for 
$E_s=150$~GeV (shown in gray), the corresponding value is 1.03. The shifts in the datasets are amplified by the $E_e^3$ factor in the ordinate. 
The HESS and Fermi error bars have been expanded to approximately include 
systematic uncertainties (apart from the energy scale uncertainties).}
\label{fig:anni}
\end{center}
\end{figure*}

We do not consider $\gamma-$ray and radio observations of the galactic center (GC) and galactic ridge since details of the DM profile are 
crucial for their interpretation.
We also do not include $\gamma-$ray observations of dwarf spheroidal galaxies which are thought to have significant concentrations of dark matter.
Purported evidence of this is that such satellite galaxies have survived several traverses of our galactic disk, stripping DM with each pass (which 
makes the use of conventional halo profiles difficult to justify).
Interesting constraints have been obtained~\cite{Bertone,meade} using HESS observations above 250~GeV of the Sagittarius Dwarf~\cite{dwarf}, and of the
GC~\cite{hessgc} and galactic ridge~\cite{hessgr}. 
For example, DM annihilation into dominantly $\tau^+\tau^-$ is disfavored as an explanation of the 
PAMELA and Fermi data for cusped halo profiles because of the accompanying large $\gamma-$ray flux from the decays of 
neutral pions~\cite{meade}. However, DM decay into only $\tau^+\tau^-$ is still viable independently of the halo profile~\cite{meade}. 
We mention this in anticipation of our results. We find that to be fully consistent with the Fermi and HESS $e^++e^-$ flux  
while simultaneously explaining PAMELA data, a somewhat 
softer spectrum than obtained from DM annihilation or decay into $e^+e^-$ is required, as can be provided by the $\mu^+\mu^-$ and $\tau^+\tau^-$ modes.   

{\it (i) Annihilating Dark Matter.}
The injection spectrum of high energy diffuse photons is
\be
\frac{d\Phi_{\gamma}}{dE_\gamma } =\frac{1}{2} \text{BF} \frac{\rho^2}{M^2_{DM}}\langle \sigma v\rangle  {d N_{\gamma}\over d E_\gamma}+c \int \sigma_{\text{\scriptsize IC}} (E_{\gamma}',E_{e}'; E_{\gamma}) n_{e}(E'_{e})
 n^{\text{\scriptsize ISRF}}_{\gamma}(E_{\gamma}')dE'_\gamma dE_{e}'\,,
\label{eq:gammas}
\ee
where the first term is the direct contribution from the annihilation process and the second term is the contribution from IC scattering of DM electrons 
(and positrons) with the ISRF at the same location. Here, $\Phi_\gamma$ is the photon flux per unit volume, BF is the boost factor assuming the 
canonical value $\langle \sigma  v\rangle = 3\times 10^{-26}$ cm$^3$s$^{-1}$,
 ${dN_\gamma \over dE_\gamma}$ is the normalized photon spectrum from each annihilation event, $n_{\gamma/e}$ is the 
interstellar photon and DM electron (and positron) densities, and $\sigma_{\text{IC}}(E_\gamma',E_e';E_\gamma)$ is the differential 
IC scattering cross section for electrons of energy $E_e'$ scattering off photons in the ISRF with energy $E_\gamma'$ and producing a photon with 
energy $E_\gamma$. A generic feature of annihilating DM is that the prompt photon flux depends sensitively on the DM halo distribution 
 as $\rho^2$.

{\it (ii) Decaying Dark Matter.}
In this case, the flux of prompt photons,
\be
\frac{1}{T}  {\rho\over M_{DM}} {d N_{\gamma}\over d E_\gamma},
\ee
replaces the first term in Eq.~(\ref{eq:gammas}) where $T$ is the DM lifetime. Since the prompt photon flux depends linearly on the DM density, 
fewer photons originate from the GC than in the annihilation scenario. 

{\it (iii) Pulsars.}
Simulations indicate that the primary $e^\pm$ and $\gamma-$ray flux from pulsars have a luminosity ratio 
$dN_{e^\pm}/dE_{e^\pm} : dN_\gamma/dE_\gamma$ between 0.2 and 0.5~\cite{Daugherty:1995zy}
 for $\gamma-$rays detected by Fermi LAT. The primary $\gamma-$ray flux is a negligible component of the diffuse flux that is dominated by
background contributions from IC scattering and $\pi^0$ decay. We include the contribution 
 from IC scattering of the primary $e^\pm$ flux emitted by pulsars.

{\bf Results.}
\label{results}
We simulate the following energy spectra, and perform a likelihood analysis as described in the appendix:
\be 
f = \left\lbrace \begin{array} {l l }
	\frac{\Phi_{e^+}}{\Phi_{e^+}+\Phi_{e^-}}\,	&	\text{for PAMELA}\\
	E^3_e{(\Phi_{e^+}+\Phi_{e^-})}\, 	&	\text{for Fermi $e^\pm$ and HESS}\\
	E^2_\gamma{\Phi_{\gamma}}\, 	&	\text{for Fermi $\gamma$-ray}\\
	\end{array}\right.
\label{eq:fs}
\ee
where $\Phi$ is the background plus signal differential flux of each species.

\begin{figure*}[b]
\begin{center}
\includegraphics[scale=0.57]{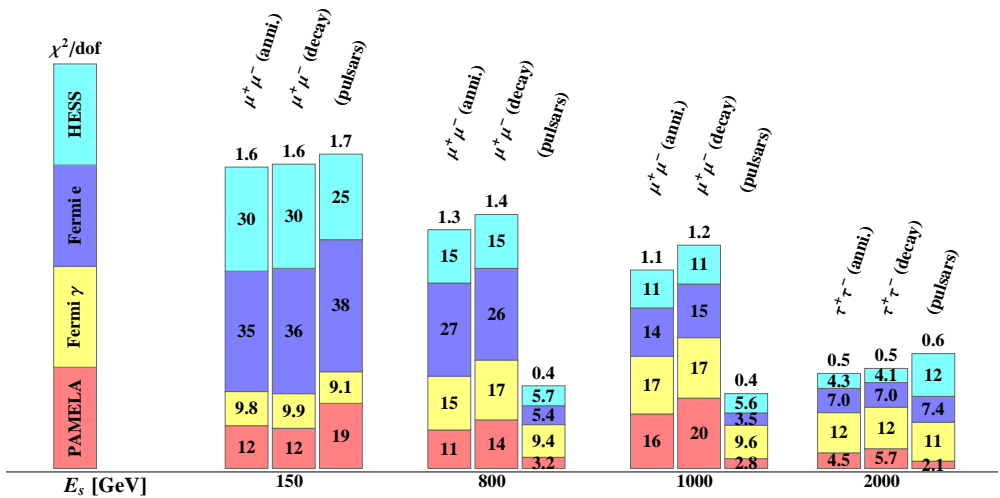}
\caption{A comparison of the best-fits for DM annihilation, DM decay and pulsars.  In the DM scenarios, the $\mu^+\mu^-$ mode
is marginally preferred over the $\tau^+\tau^-$ mode, except for $E_s=2$~TeV. }
\label{fig:barchart}
\end{center}
\end{figure*}

\begin{figure*}[t]
\begin{center}
\includegraphics[scale=0.7]{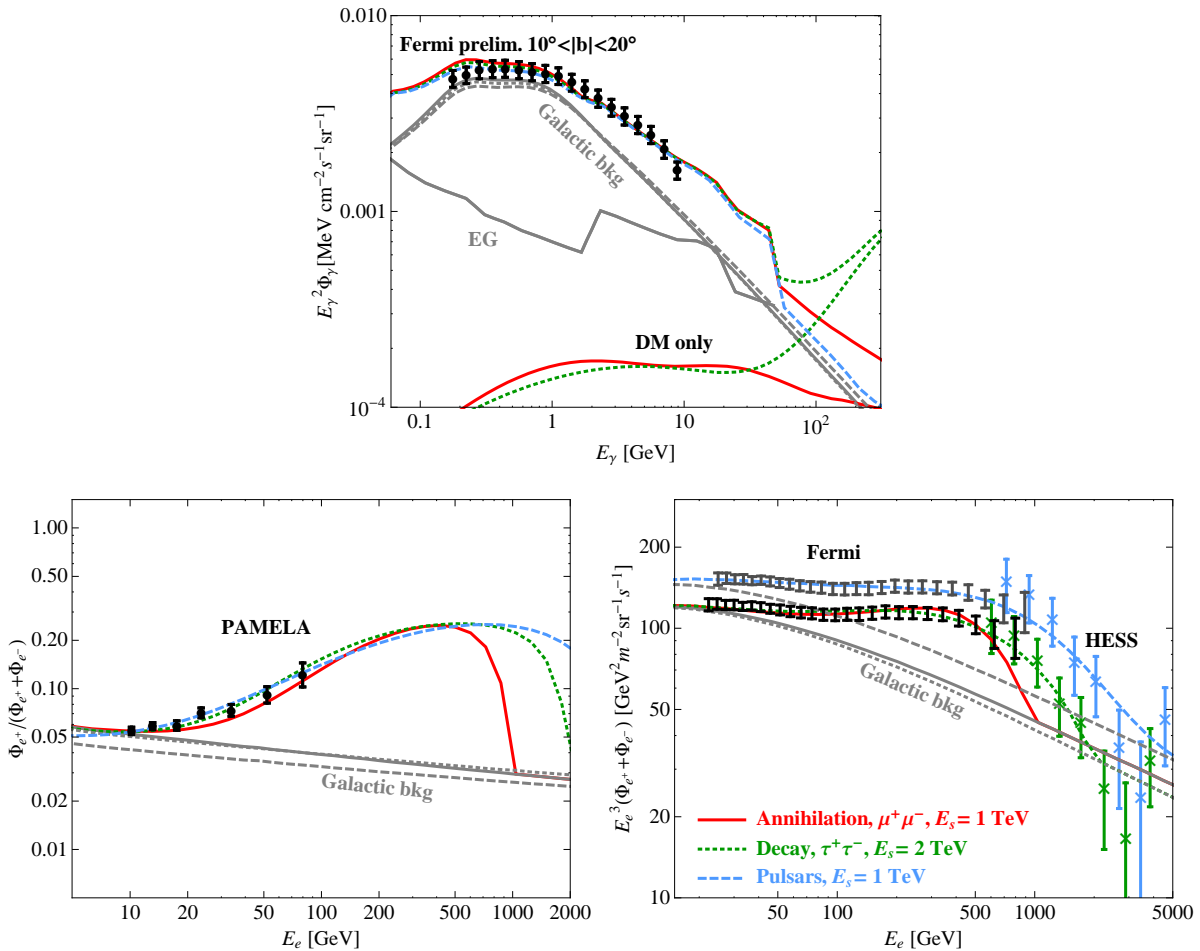}
\caption{Illustrative spectra for DM annihilation into $\mu^+\mu^-$ ($E_s=1$~TeV), DM decay into $\tau^+\tau^-$ ($E_s=2$~TeV), and pulsars with $E_s=1$~TeV. 
Note that the $e^\pm$ spectra for DM annihilation into $\tau^+\tau^-$ with $E_s=2$~TeV are identical to the DM decay case. 
The pulsar-only contribution to $\gamma-$rays lies slightly outside the range of the figure. 
The corresponding $\chi^2$ values can be found in Fig.~\ref{fig:barchart}.
For the DM decay case, the green HESS data points are systematically shifted to lower energies with $\epsilon=0.84$. 
For pulsars, the light blue HESS data points have $\epsilon=1.0$. The HESS data 
corresponding to DM annihilation have $\epsilon=0.8$, and are not shown. 
For the DM cases, the Fermi $e^\pm$ data shift (shown in black) with an energy calibration scale of 0.9, and for pulsars (shown
in gray), the corresponding value is 1.0. The electron backgrounds have $\gamma_0 \simeq 2.61$.
}
\label{fig:adp}
\end{center}
\end{figure*}

In Fig.~\ref{fig:baranni} we show the best-fit $\chi^2$ values for DM annihilation into various charged lepton final states. 
The combined analysis includes the Fermi LAT, 
PAMELA, and HESS datasets. For PAMELA, we only analyze data above 10~GeV for which the solar modulation effects are negligible. 
We do not include WMAP haze data in our $\chi^2-$analysis because it may be subject to unknown and potentially large systematic uncertainties. 
It is evident that the data prefer softer spectra. 
The Fermi $e^\pm$ data and steepening of the HESS spectrum above 1~TeV are easily reproduced by heavy DM ($M_{DM}=2$~TeV) annihilating to 
the $\tau^+\tau^-$ mode; see Fig.~\ref{fig:anni}. The results for DM decay are only marginally worse, so we do not show them.

Figure~\ref{fig:barchart} shows a comparison of our three scenarios, all of which satisfactorily explain the experimental data.
DM annihilation explains the data marginally better than DM decay. For DM annihilation, the $\rho^2$ enhancement near the GC yields a larger 
fraction of $e^\pm$ that propagate a long distance and hence a slightly softer $e^\pm$ spectrum that fits PAMELA and Fermi $e^\pm$ data better. 
For DM annihilation and decay into $\mu^+\mu^-$, we find the best-fit boost factor and lifetime to satisfy the approximate relations,
\be
BF = 431 E_s - 38.9\,, \quad \quad\quad
{T \over 10^{26} \, \rm{sec}}= 2.29 + {0.591\over E_s}\,,
\label{bf}              
\ee
with $E_s$ in TeV. The exponential cut off in the pulsar electron spectrum can reproduce the HESS data very well. 
Notice that the the quality of the fit is only weakly dependent on $E_s$ for pulsars.  

In Fig.~\ref{fig:adp}, we show representative best-fit spectra from the joint analyses of the Fermi, PAMELA, and HESS data.  
The mid-latitude preliminary Fermi $\gamma-$ray data reside in the energy range where the galactic background significantly dominates the DM contribution.
Above about 50~GeV, the DM signals dominate, and may be resolvable with future Fermi LAT data up to 300~GeV.
Even with future data, the isotropic EG component is significant at mid-latitudes and needs to be carefully investigated 
before robust conclusions can be drawn.
From Fig.~\ref{fig:adp} we see that the $\gamma-$ray contribution from pulsars is negligible.

\begin{figure*}
\begin{center}
\includegraphics[scale=0.7]{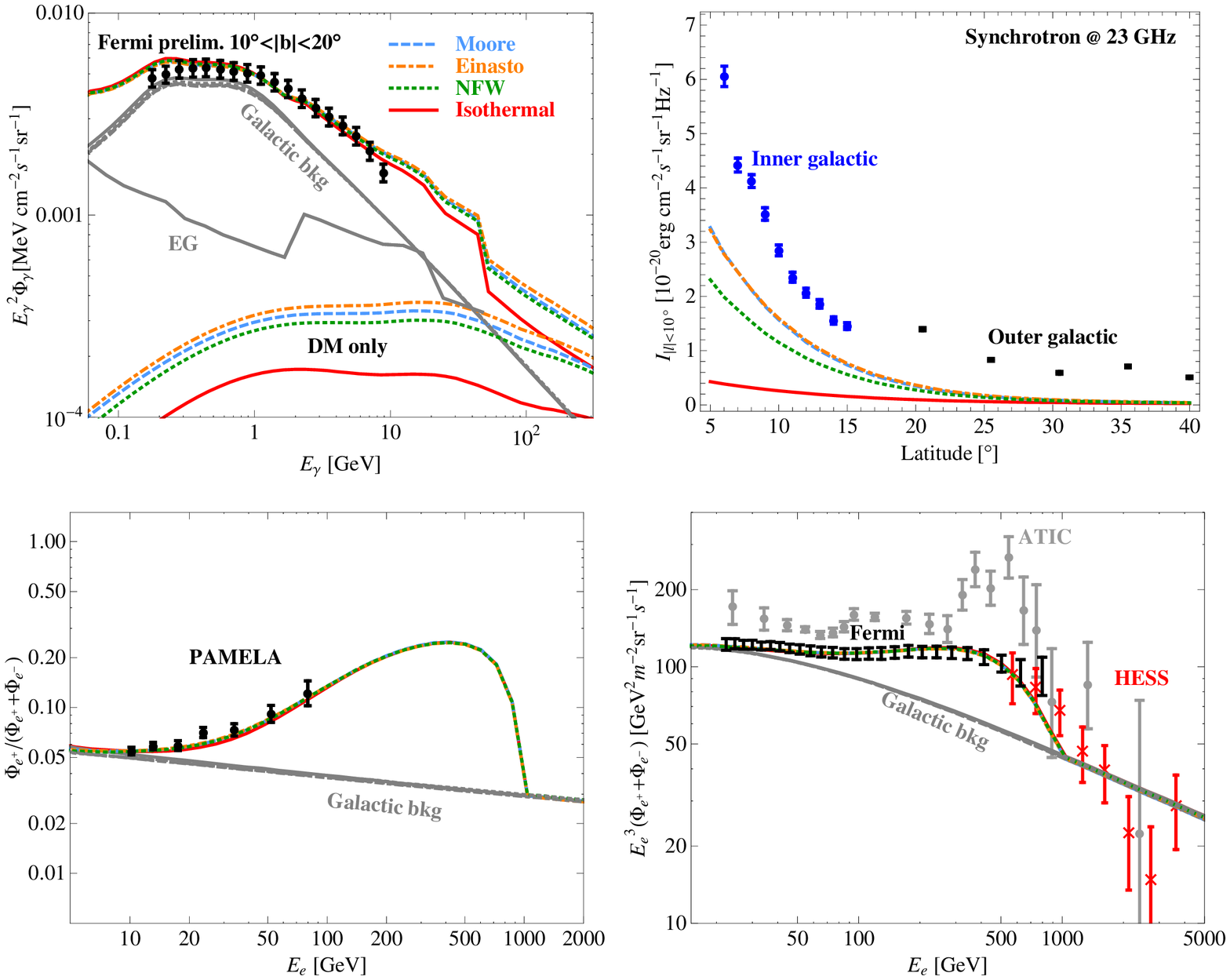}
\includegraphics[scale=0.57]{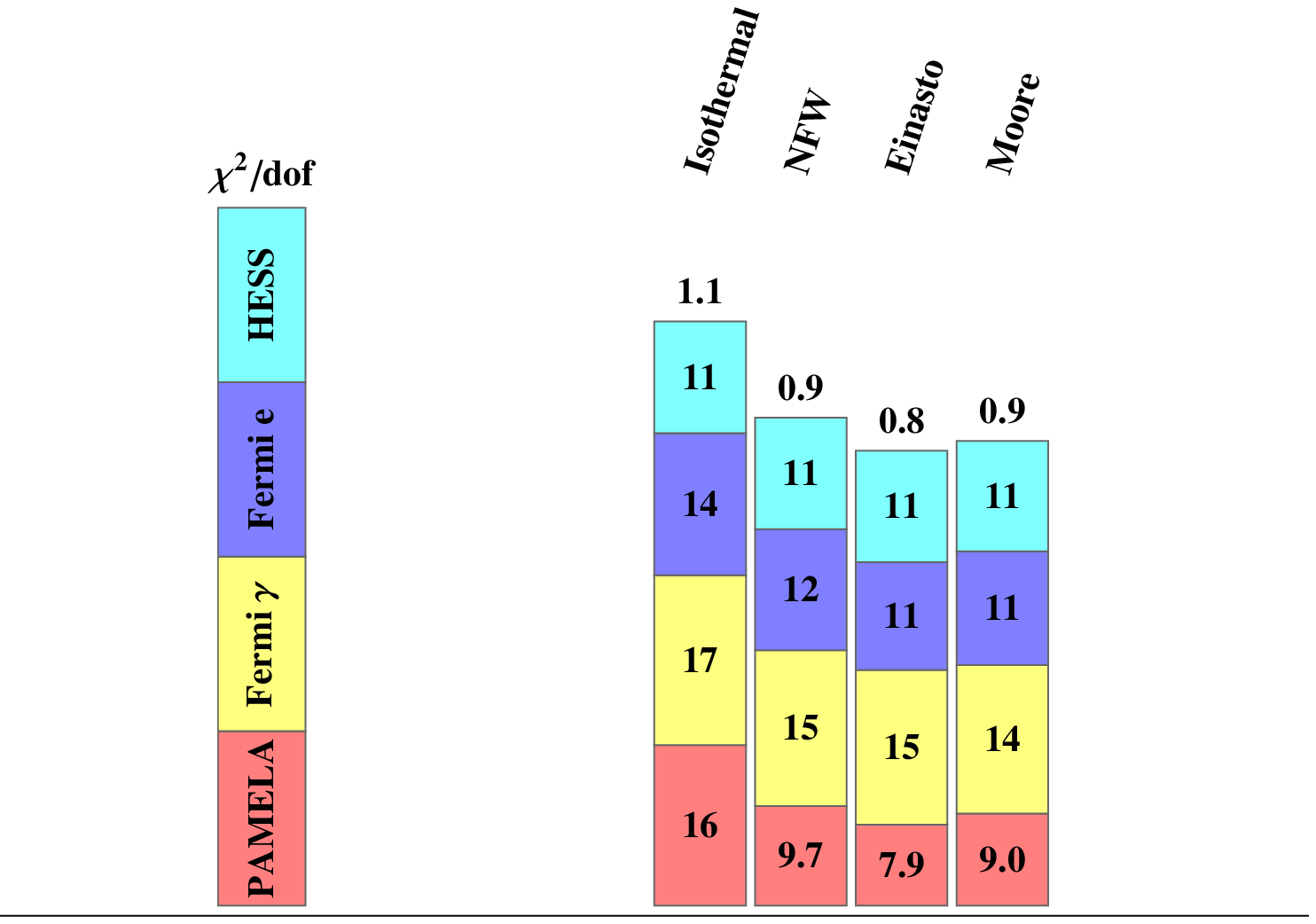}
\caption{Impact of different DM halo distributions for DM annihilation into $\mu^+\mu^-$ with $E_s=1$~TeV.  The cuspy Moore, Einasto and NFW profiles 
yield higher contributions to the diffuse $\gamma$ and synchrotron spectra than the isothermal profile. 
The $e^{\pm}$ spectra for the different profiles overlap and have similar boost factors. The HESS and Fermi energy calibration scales are 
$\epsilon=0.8$ and 0.9, respectively, and are the same for all profiles. 
The WMAP haze data from the outer~\cite{Cumberbatch:2009ji} and inner~\cite{finkbeiner2} galactic regions are shown at a representative frequency and
only with statistical uncertainties, are not included in the $\chi^2-$analyses, and are unexplained by our scenarios. 
The ATIC data are not included in the fit, but are displayed for comparison.
The corresponding $\chi^2$ values show that the data do not favor a particular profile.
}
\label{fig:profile}
\end{center}
\end{figure*}

In Fig.~\ref{fig:profile}, we show how DM annihilation signals depend on the DM halo profile. 
We single out the annihilation case because of its stronger dependence on $\rho$. For a cuspier profile, 
a larger fraction of $e^{\pm}$ are injected near the GC so that more synchrotron emission occurs as they propagate through the galaxy. 
This is evident
in the upper right-hand panel of  Fig.~\ref{fig:profile}. We see that there is a marginal preference for
cusped profiles essentially because the combination of PAMELA and Fermi $e^{\pm}$ data require a not-too-hard $e^{\pm}$ spectrum. 
The different profiles produce similar $e^{\pm}$ spectra, and require boost factors of about 400 with a difference of less than $3\%$ among them.

\begin{figure*}[t]
\begin{center}
\includegraphics[scale=0.62]{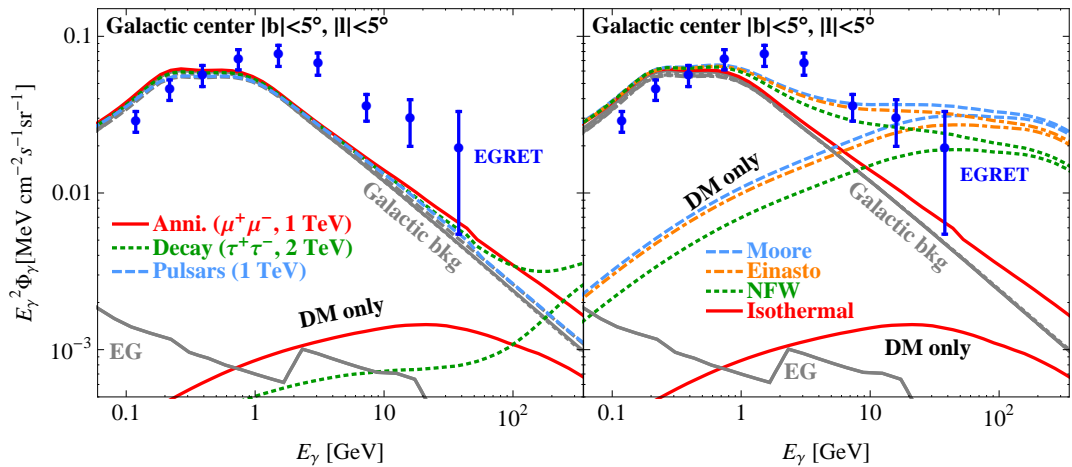}
\caption{Diffuse gamma-ray spectra in the GC region $|l|,|b|<5^\circ$. The left-hand panel shows spectra corresponding
to the DM annihilation, decay and pulsar cases in Fig.~\ref{fig:adp}, where DM assumes the
isothermal halo profile. The right-hand panel shows the spectra corresponding to the
DM annihilation cases with different profiles in Fig.~\ref{fig:profile}. 
EGRET data~\cite{Strong:2005zx} are plotted to illustrate that a significant diffuse
gamma-ray excess can be expected from cuspy profiles in the GC region
although preliminary Fermi LAT results see no excess in the mid-latitude range.
}\label{fig:central}
\end{center}
\end{figure*}

Although Fermi LAT does not corroborate the $\gamma-$ray excess in EGRET data at mid-latitudes, it is still quite possible that it may observe an excess
at the GC; see Fig.~\ref{fig:central}.  In this region of the sky the DM signal dominates the EG background. 
In contrast with the isothermal profile, the cusped profiles enhance the flux in this energy range, signaling DM annihilation and
a cusped halo profile.

{\bf Summary.}
Three popular explanations have been proposed for the sharp rise in the positron fraction in PAMELA data: 
DM annihilation, DM decay and pulsar sources. Associated with these possibilities are $\gamma-$ray signals. We deliberately avoided
existing observations of the galactic center since they are sensitive to the details of the halo profile, and focused on the diffuse $\gamma-$ray 
background from latitudes above $10^\circ$. Preliminary
results from the Fermi LAT do not confirm the diffuse $\gamma-$ray excess at mid-latitudes reported by the EGRET observatory. 
With these data that are consistent with the expected background, we entertained the possibility that 
final state radiation and inverse Compton scattering (which are the primary sources of $\gamma-$rays) may help constrain DM as an explanation
of the PAMELA, Fermi $e^\pm$ and HESS data since these data require copious production of positrons and electrons. However, 
we found that this is not the case for DM annihilation/decay into charged leptons; our choice of leptophilic annihilation/decay
was dictated by the nonobservance of an excess in the $\bar{p}/p$ flux ratio measured by PAMELA. 
We emphasize that the preliminary Fermi results are in a $\gamma$ energy range ($<10$~GeV) where the galactic and  
extragalactic backgrounds  overwhelm the DM contribution. With upcoming observations up to 300~GeV, Fermi data promises to 
be far more constraining.
Dark matter with $E_s$ between 800~GeV and 2~TeV 
is easily consistent with all 4 datasets for both annihilation and decay. The data prefer the softer spectra that arise from 
the $\mu^+\mu^-$ and $\tau^+\tau^-$ modes; see Fig.~\ref{fig:barchart}.
The boost factors required for DM annihilation and lifetimes for
DM decay follow Eq.~(\ref{bf}) for the $\mu^+\mu^-$ mode. Overall, pulsars have the least tension with the data, primarily because
the HESS data show a gradual fall-off in the energy spectrum, which is characteristic of pulsar sources. To draw these conclusions we varied
the backgrounds within reason, accounted for available systematic uncertainties, and assumed an isothermal profile. It is no surprise
that with an isothermal profile, the WMAP haze could not be accommodated.

For DM annihilation, the source term depends on the DM density as $\rho^2$, so effects of the halo distribution are especially relevant. 
We considered 4 different profiles and found that the 4 datasets when analyzed jointly 
show no appreciable preference for the degree of cuspiness.
The observance of a significant gamma-ray excess from the galactic center but no excess at mid-latitudes is a striking signal
of DM annihilation and a cusped DM halo.


\vspace{0.2cm}
{\it Acknowledgments. }
We thank M.~Medvedev for a discussion. This work was supported by DOE Grant Nos. DE-FG02-04ER41308, DEFG02-95ER40896, DE-FG02-84ER40173 and DE-AC02-06CH11357, by NSF Grant No. PHY-0544278, and by the Wisconsin Alumni Research Foundation.


\vspace{0.2cm}

{\bf Appendix.}
\underline{\it {Numerical procedure.}}
To fully simulate the $e^\pm$ and photon spectra, we use the GALPROP package~\cite{Strong:1999sv} with the ISRF profile~\cite{Porter:2005qx}.  
The DM $\gamma-$ray and $e^\pm$ injection spectra are calculated using DMFIT~\cite{Jeltema:2008hf} and MicrOMEGAs~\cite{Belanger:2008sj}, respectively. 
We run GALPROP in the 2-dimensional mode and use the ``conventional" (ID:500180) model~\cite{Bergstrom:2004cy} for the galactic set-up and cosmic 
ray source/injection parameters; however, we vary the injection spectral index $\gamma_0$ for the electron background (parameterized as a power-law
$E^{-\gamma_0}$) in the range $2.2-2.9$. 
This model produces a $\bar{p}/p$ spectrum that agrees with PAMELA data above 10~GeV. The  
cylindrical diffusion zone has a radius of 20~kpc and a height of 4~kpc; the magnetic field that determines the synchrotron radiation intensity is given by 
\be 
B=B_\odot \text{e}^{-\frac{{\tt r}-{\tt r}_\odot}{{\tt r}_s}} \text{e}^{-\frac{|z|}{z_s}}\,,
\ee
with a default local strength $B_\odot=5$~$\mu$G and scales ${\tt r}_s=10$~kpc, $z_s=2$~kpc. 
The final $\gamma$ spectra are calculated from the GALPROP created sky-maps by averaging the IC, FSR, bremsstrahlung, and $\pi^0$ decay 
contributions over specific ranges of galactic coordinates. 
The galactic and source spectra are generated separately and then combined with a varying normalization factor for the latter.



Cosmic ray propagation is governed by the energy loss - diffusion equation,
\be 
\frac{d\Phi}{dt}-D(E)\cdot \bigtriangledown^2\Phi-\partial_{E}(D_p(E)\cdot \Phi)=Q\,,
\ee
where $Q$ is the injection source term for each species. The energy loss coefficient $D_p(E)$ is dynamically evaluated and for
 $e^{\pm}$ it is mainly determined by bremsstrahlung and IC scattering. The spatial diffusion parameter $D(E)$ is parametrized~\cite{Strong:2007nh} as
\be
D(E) = \beta D_0 \left({R(E) \over R_0 }\right)^\delta \text{cm}^2\text{s}^{-1}\,,
\ee
where $\beta=v/c=1$ for cosmic rays of interest.  $R(E)$ and $R_0$ are the particle rigidity at energy $E$ and the reference rigidity, respectively. 
For $e^\pm$, $R(E)=E$. To account for the uncertainty in cosmic ray diffusion, we vary the diffusion parameters ($D_0, R_0, \delta$). 
Since the various $D(1\,\rm{GeV})$ fall in the range $(3-5)\times10^{28}$~cm$^2$/s to fit cosmic ray nuclei data~\cite{Strong:2007nh}, we
impose this as a constraint in our analysis.
We restrict $\delta$ to be above $1/3$, the value expected for a turbulent magnetic field with a Kolmogorov spectrum.
For each $E_s$, we also vary the total background $e^-$ normalization, 
and the BF (for DM annihilation), or $T$ (for DM decay), or $N$ (in Eq.~\ref{pulsardist} for pulsars).
 

For the DM halo, we considered the following halo distributions:
\begin{equation}
\rho(r)=\left\lbrace \begin{array}{ll}
\rho_{\odot}\frac{1+(r_\odot/r_s)^2}{1+(r/r_s)^2} & \text{Isothermal, $r_s=5$~kpc ~\cite{IsoT}}\\
\rho_{0}\, \frac{r_s}{r}\left(1+\frac{r}{r_s}\right)^{-2} & \text{NFW, $\rho_0=0.26 \text{ GeV}/\text{cm}^3$, $r_s=20$~kpc~\cite{Navarro:1995iw}}\\
\rho_{\odot}\text{exp}\lbrace-\frac{2}{\alpha}[(r^{\alpha}-r_{\odot}^\alpha)/r_s^\alpha]\rbrace & \text{Einasto, $\alpha=1.7$, $r_s=25$~kpc ~\cite{Navarro:2008kc}}\\
{\rho_\odot}\left(\frac{r_{\odot}}{r}\right)^{1.16}\left(\frac{1+r_\odot/r_s}{1+r/r_s}\right)^{1.84} & \text{Moore, $r_s=30$~kpc ~\cite{Diemand:2005wv}}\\
\end{array}\right.
\end{equation}
where the local DM density is $\rho_{\odot}=0.3$~GeV/cm$^3$ for all profiles, and $r_s$ is the size of the DM halo.
The spatial source of $e^\pm, \gamma$ is obtained by convolving the halo distribution with the injection spectra.

{\underline{\it{Likelihood analysis.}}
We consider Fermi, PAMELA and HESS as independent sets and sum the $\chi^2$ contribution from each set. In obvious notation,
\be
\chi^2=\sum_{experiments}\sum_{i}\frac{(f^{th}_i-f^{ex}_i)^2}{(\Delta f^{ex}_i)^2}\,,
\ee
where $f$ are the spectra in Eq.~(\ref{eq:fs}).
Since bin sizes are available for PAMELA and Fermi $e^\pm$ data, we average over each bin to obtain the theoretical expectation $f^{th}$.

The HESS data have a $15\%$ energy calibration uncertainty. To account for it we fit an additional parameter 
$\epsilon=1+\frac{\Delta E}{E}$ that
 scales a data point from ($E, E^3\frac{dN}{dE}$) to ($\epsilon E, \epsilon^2 E^3\frac{dN}{dE}$), while
the number count $E\frac{dN}{dE}$ remains fixed.
With $f(E)=E^3\frac{dN}{dE}$,
\be 
\chi^2_{HESS}(\epsilon)=\sum_{i}\frac{(f^{th}(\epsilon E_i)-\epsilon^2f^{HESS}_i)^2}{(\epsilon^2\Delta f^{HESS}_i)^2}+\frac{(1-\epsilon)^2}{(\Delta\epsilon)^2}\,,
\label{eq:HESS}
\ee
where $\Delta\epsilon=0.15$ and $\Delta f^{HESS}_i$ includes the additional systematic uncertainty (from modeling hadronic interactions and the atmosphere) 
in quadrature with the statistical uncertainty~\cite{Collaboration:2008aaa}. The Fermi $e^\pm$ data have a $^{+5}_{-10}$\% energy calibration uncertainty
which we treat in analogy with the HESS calibration uncertainty, by including an additional energy scale parameter, 
except for one important difference: We allow this parameter to vary only within the $1\sigma$ range of the calibration uncertainty so that the Fermi
$e^\pm$ data do not rigidly shift too much and become discordant with data from AMS, HEAT, ATIC, BETS and PPB-BETS which are consistent with each other 
between 10~GeV and 100~GeV. We have checked that the Fermi dataset shifts
 significantly unless low energy data from other experiments are included in the analysis; the low energy data from ATIC, for example, anchor the 
Fermi energy calibration so that the scale parameter does not deviate too far from unity.

\end{document}